\documentclass{article}
\usepackage{PSIP2003,amsmath,amsfonts,amssymb,,epsfig,url,moreverb}

\newcommand{\E}{\mathsf{I\!E}}
\newcommand{\R}{\mathbb{R}}
\newcommand{\Z}{\mathbb{Z}}

\newcommand{\abs}[1]{\lvert#1\rvert}

\title{Testing the normality of the gravitational wave data with a low cost recursive estimate of the kurtosis}
\name{E. Chassande-Mottin\thanks{The author thanks the European Gravitational Observatory for its hospitality.}}
\address{ILGA-Virgo (CNRS -- UMR6162), Observatoire de la C\^ote d'Azur\\ 
BP 4229 F-06304 Nice Cedex 4 France}

\begin{document}
%\ninept
%
\maketitle
\begin{abstract}
  We propose a monitoring indicator of the normality of the output of a gravitational wave detector.
  This indicator is based on the estimation of the kurtosis (i.e., the 4th order statistical moment
  normalized by the variance squared) of the data selected in a time sliding window. We show how a
  low cost (because recursive) implementation of such estimation is possible and we illustrate the
  validity of the presented approach with a few examples using simulated random noises.
\end{abstract}

\vspace{2mm}
Four large-scale detectors \cite{gwdetectors} of gravitational waves (GWs) are on the point to take
their first scientific data. They all rely on the principle used in the Michelson experiment:
measure the relative length $\delta L/L$ of the two perpendicular arms (each formed by two suspended
masses) of the detector. The goal is to reach a measurement sensitivity (i.e., decrease the noise
level) such that the small changes of $\delta L$ caused by the GWs emitted from astrophysical
sources (such as the coalescing binaries of neutron stars or black holes) can be detected when they
pass through the instrument. 

Since only a small number of such events are likely to be observed, the problem consists from the
data analysis viewpoint, in looking for rare transients appearing in the detector output. For the
coalescing binaries mentioned above, this will be done by implementing a bank of matched filters:
each filter correlate the data with the expected GW emitted from a binary and we repeat this
operation for a large number of possible targets. Assuming the template waveform are reliable, the
matched filtering approach can be shown to be optimal in the Neymann-Pearson sense provided that the
noise is Gaussian.  Therefore, it is crucial to monitor the noise Gaussianity and locate any
departures from this nominal hypothesis.

There are many ways for the noise to departs from Gaussianity, however not all of them are relevant
for the problem of GW detection. One of the possibilities is to have a noise probability density
function (PDF) with heavier tails than the Gaussian bell curve. This particular discrepancy is a
problem because it causes a increase of the false detection rate (i.e., the large values in the
tails make the matched filter triggers more often).

The kurtosis \cite{johnson70:_distr_statis} is a well-known measurement of the decay rate of the PDF
in the tails. This motivates us to use the kurtosis as an index measuring normality. We define
the mean of the signal $x(t)$ by $\mu_1(t)\equiv\E[x(t)]$ where $\E[\cdot]$ is the expectation
operator.  The central moments \cite{johnson70:_distr_statis} of $x(t)$ are given by
$\mu_r(t)\equiv\E[(x(t)-\mu_1(t))^r]$ where $r\geq 2$ is the order.  The kurtosis is defined by the
following ratio $\kappa_4(t)\equiv\mu_4(t)/(\mu_2(t))^2$.

Because the analysis must be done in real-time, the normality test should not be computationally
expensive. We propose here an efficient (because recursive) implementation of the kurtosis
estimation in a short-term and sliding observation window. Another recursive estimator of the
kurtosis was proposed in \cite{amblard95:_adapt} and required to have zero-mean signals. The
presented approach here works also with non-centered signals.

The outline of the paper is as follows. We choose a simple mathematical structure for the short-term
and recursive estimation of the central statistical moments. In Sect. \ref{recursive_est}, we
identify in the selected family which estimators of $\mu_1$, $\mu_2$ and $\mu_4$ have a vanishing
bias. We then show in Sect. \ref{kurtosis_est} that an adequate Taylor approximation of the ratio of
the unbiased estimates of $\mu_4$ and $\mu_2$ obtained previously yields the recursive estimate of
$\kappa_4$ and we detail its computation algorithm. Finally, we apply in Sect. \ref{validation} the
proposed estimator to a few illustrative cases and we explain how it is used in practice for the
monitoring of the normality of the GW detector output.

\section{Recursive estimates with vanishing bias}
\label{recursive_est}
\subsection{Recursive estimate of the mean and variance}
We assume that the signal $x(t),\:t\in\Z$ (using a unit sampling rate) is locally stationary (i.e.,
its statistical moments do not change during a finite time period $\delta t_{stat}$) and ergodic
(i.e., its statistical moments can be estimated from its samples). The mean of $x(t)$ can be
estimated with the following weighted average of the data selected by the window $w_1(t)$
\begin{equation}
\label{loc_av_mu1}
\hat{\mu}_1(t)\equiv C_{1}\sum_{n=0}^{t} w_1(t-n)\:x(n),
\end{equation}
where the duration of $w_1(t)$ is smaller than $\delta t_{stat}$. If the window in use is of
exponential type, i.e.  $w_1(t)\equiv a_1^t$, this estimator can be equivalently calculated
recursively with
\begin{equation}
\label{rec_mu1}
\hat{\mu}_1(t)= a_1 \hat{\mu}_1(t-1) + C_1 x(t).
\end{equation}

The problem is to find the constants $a_1>0$ and $C_1$ such that $\hat{\mu}_1(t)$ is asymptotically
unbiased i.e., $\E[\hat{\mu}_1(t)]=\mu_1$ when $t \rightarrow +\infty$. The bias can be calculated
directly from the definition of the estimator, yielding
\begin{equation*}
\E[\hat{\mu}_1(t)]= C_{1}\mu_1 \frac{1-a_1^{t+1}}{1-a_1} \rightarrow \frac{C_1}{1-a_1}\mu_1 \quad \text{when $t \rightarrow +\infty$}
\end{equation*}
and assuming $a_1<1$. We conclude that $\hat{\mu}_1(t)$ is an unbiased estimator of the mean if
$C_1=1-a_1$. 

The same method can be applied to find recursive and unbiased estimators of the higher order
moments, based on the following choice of expression~:
\begin{equation}
\label{def}
\hat{\mu}_r(t)\equiv C_r \sum_{n=0}^{t} w_r(t-n)\:(x(n)-\hat{\mu}_1(n-1))^r,
\end{equation}
where $r\geq 2$. Similarly to the mean, a recursive implementation is possible if we choose
$w_r(t)\equiv a_r^t$. We restrict to the interval $0<a_r<1$.

When $r=2$, we essentially average the squared differences to the estimated mean in a sliding time
window defined by $w_2(t)$.  We evaluate the bias of $\hat{\mu}_2(t)$ by taking the expectation of
(\ref{def}). We first evaluate~:
\begin{multline*}
\E[(x(n)-\hat{\mu}_1(n-1))^2]=\mu_1^2+\mu_2-2C_{1}\mu_1^2 \frac{1-a_1^n}{1-a_1}\\
+C_1^2\left(\mu_2\frac{1-a_1^{2n}}{1-a_1^2}+\mu_1^2\left(\frac{1-a_1^n}{1-a_1}\right)^2\right),
\end{multline*}
and setting $C_1=1-a_1$ (i.e., set the bias of $\hat{\mu}_1$ to zero), the summation leads to~:
\begin{multline*}
  \E[\hat{\mu}_2(t)]= \frac{2 C_2 \mu_2}{1+a_1}\:\frac{1-a_2^{t+1}}{1-a_2} +C_2\left(\mu_1^2-\frac{1-a_1}{1+a_1}\mu_2\right)\\
\times \frac{a_2^{t+1}-a_1^{2(t+1)}}{a_2-a_1^2} \rightarrow \frac{2 C_2}{(1+a_1)(1-a_2)}\mu_2,
\end{multline*}
when $t\rightarrow +\infty$ provided that $a_1$ and $a_2<1$. Consequently, the bias of
$\hat{\mu}_2(t)$ is zero when $C_2=(1+a_1)(1-a_2)/2$.

\subsection{Extension to the 4th order}
We proceed to the fourth order as previously, starting from the definition (\ref{def}) with $r=4$.
The evaluation of the bias of $\hat{\mu}_4(t)$ requires the calculation of
\begin{multline*}
  \E[(x(n)-\hat{\mu}_1(n-1))^4]=(1+f_4)\mu_4\\
  +4(1-f_1-f_3+ f_1 f_3)\mu_1\mu_3+ 3(2f_2+f_2^2-f_4)\mu_2^2\\
  +6(1-2 f_1 + f_2+ f_1^2 - 2 f_1 f_2+ f_1^2 f_2)\mu_1^2\mu_2\\
  +(1-4 f_1 +6 f_1^2-4 f_1^3 + f_4)\mu_1^4,
\end{multline*}
where we defined $f_k\equiv C_1^k(1-a_1^{kn})/(1-a_1^k)$.

Setting $C_1=1-a_1$ so that the estimate of the mean is not biased, and making the summation for
$n=0\ldots t$, we get
\begin{equation}
\label{bias4}
\E[\hat{\mu}_4(t)]=\frac{C_4}{1-a_4}(k_4 \mu_4+k_2 \mu_2^2) + r(t),
\end{equation}
where $r(t)$ is a complicated sum of integer powers of terms of the form $K a_1^{(t+1)}$ and $K
a_4^{(t+1)}$ where $K \in \R$. The constants can be expressed as
\begin{align*}
k_4 &= \frac{2(1-a_1+2a_1^2)}{(1+a_1)(1+a_1^2)}&
k_2 &= \frac{6(1-a_1)(1+a_1+2a_1^2)}{(1+a_1)^2(1+a_1^2)}.
\end{align*}

It is interesting to note that the first term in (\ref{bias4}) does not depend on $\mu_3$.

% \begin{align*}
% k_4 &=1+\frac{C_1^4}{1-a_1^4}&
% k_2 &=6\frac{C_1}{1-a_1^2}+3\frac{C_1^4}{(1-a_1^2)^2}-3\frac{C_1^4}{1-a_1^4}
% \end{align*}
If $a_1$ and $a_4<1$, the function $r(t)$ goes to 0 when $t$ tends to $+\infty$ so that the
expectation of $\hat{\mu}_4(t)$ depends in a simple manner of $\mu_4$ (the objective value) and
$\mu_2^2$.  Assuming that $\mu_2$ is known, we can set the bias of $\hat{\mu}_4(t)$ to a simple
constant offset if we choose the coefficient of $\mu_4$ in (\ref{bias4}) equal to 1, i.e.
we set $C_4=(1-a_4)/k_4$ in the following.

\section{Taylor approximation of the kurtosis estimator}
\label{kurtosis_est}
The results of the previous Section motivate us to propose the following estimator
$\hat{\kappa}_4(t)$ of the kurtosis, obtained by dividing (\ref{bias4}) by $\mu_2^2$ and replacing
the variance by its recursive estimate~:
\begin{equation}
\label{est_k}
\hat{\kappa}_4(t)=\bar{\kappa}_4(t)-\frac{k_2}{k_4},
\end{equation}
where $\bar{\kappa}_4(t)\equiv\hat{\mu}_4(t)/\hat{\mu}^2_2(t)$ is a heuristic estimator which
we correct in (\ref{est_k}) by suppressing the offset.

We choose the same window for all the involved estimators (i.e., we set $a_4=a_2=a_1$), so that
\begin{equation*}
\bar{\kappa}_4(t)=\frac{a_1\mu_4(t-1)+C_1(x(t)-\hat{\mu_1}(t-1))^4}{(a_1\mu_2(t-1)+C_1(x(t)-\hat{\mu_1}(t-1))^2)^2}.
\end{equation*}

For sufficiently large duration of the window ($a_1 \rightarrow 1$), we can treat $C_1$ as an
epsilon and approximate this ratio to the first two terms of its Taylor expansion (this idea was
inspired by \cite{aarts02:_effic}), which leads to the following expression~:
\begin{multline}
\label{rec_k}
\bar{\kappa}_4(t) = \left(1+C_1-2C_1\delta^2 x(t)\right)\:\bar{\kappa}_4(t-1)\\
+C_1\delta^4 x(t) +O(C_1^2),
\end{multline}
where $\delta^2 x(t)\equiv(x(t)-\hat{\mu}_1(t-1))^2/\hat{\mu}_2(t-1)$ computes a normalized distance
of the current signal sample to the mean. The correct estimate of the kurtosis proposed in
(\ref{est_k}) is obtained by subtracting the offset given by
\begin{equation*}
\frac{k_2}{k_4}=\frac{-3C_1(4-5C_1+2C_1^2)}{(2-C_1)(2-3C_1+2C_1^2)}=-3C_1+O(C_1^2),
\end{equation*}
to the approximation in (\ref{rec_k}). Note that since $C_1\ll 1$, the offset can be neglected in
most practical situations because $\kappa_4 \gtrsim 1$.

The recursive estimation of the kurtosis obtained in (\ref{rec_k}) is intuitively appealing~: if we
replace the estimators of the mean and variance by their actual values in the definition of
$\delta^2 x(t)$, then noting that $\E[\delta^2 x(t)]=1$ and $\E[\delta^4 x(t)]=\kappa_4(t)$, we can
write
\begin{equation}
\bar{\kappa}_4(t) \approx a_1\:\bar{\kappa}_4(t-1)+C_1 \kappa_4(t).
\end{equation}

Using the equivalence between (\ref{loc_av_mu1}) and (\ref{rec_mu1}), we conclude that eq. (\ref{rec_k})
may be seen as a local average of the crude estimation of $\kappa_4(t)$ made by $\delta^4
x(t)$.

Eq. (\ref{rec_k}) yields a recursive estimation scheme which is described by the pseudo-code in Tab. \ref{code}

\section{Validation and practical use}
\label{validation}
In this section, we make various numerical checks of the proposed method. In all the simulations,
the estimator $\hat{\kappa}_4(t)$ as defined in eqs. (\ref{est_k}) and (\ref{rec_k}) is computed
with $C_1=2.9912 \times 10^{-3}$. This corresponds to a window duration of about 20 s if a sampling
rate $f_s=50$ Hz is assumed.

\begin{figure}
\begin{minipage}[b]{1.0\linewidth}
  \centering
  \centerline{\epsfig{figure=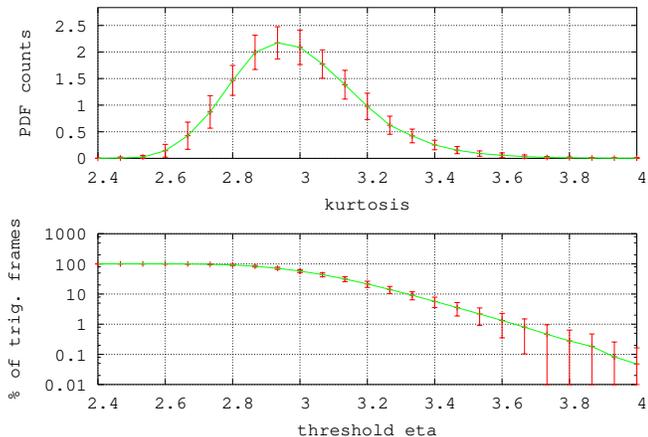,width=\textwidth}}
\end{minipage}
 \caption{\label{pdf}
   \textbf{Probability density function of $\hat{\kappa}_4$ and rate of triggered frames computed
     with Gaussian noise} {\protect \cite{02:_gnu_octav}}. The results presented in this figure were
   obtained using 50 streams of simulated (zero mean, unit variance and white) Gaussian noise.  Each
   data stream contains 30,000 samples (i.e., 600s if the sampling rate is $f_s=50$ Hz).
   \textit{top}: the empirical histogram presented in this diagram gives an estimation (bounded by
   error bars) of the PDF of $\hat{\kappa}_4(t)$ (see Sect. {\protect \ref{validation}} for
   details). \textit{bottom}: for a threshold $\eta$ taking values between $2.4$ and $4$, we present
   the percentage of frames of data where $\hat{\kappa}_4(t)>\eta$ at least once. To make this
   computation, the data was divided into 600 chunks of 1 second duration (each of them
   defining a frame) and the first 20 frames (i.e., equivalent to the window duration) of each
   trial were removed.}
\end{figure}

\noindent \textbf{Check \#1: what is the bias of the estimator? ---}
We answer this question in the case of a Gaussian noise for which $\kappa_4=3$.  Figure \ref{pdf}
(top) shows the histogram of $\hat{\kappa_4}(t)$ computed with a simulated (zero-mean, unit variance
and white) Gaussian noise. With this histogram, we evaluate the expectation of $\hat{\kappa_4}(t)$
to be equal to $3.0006$ with a bin size of $\pm 0.03$.

\begin{figure}
\begin{minipage}[b]{1.0\linewidth}
  \centering
  \centerline{\epsfig{figure=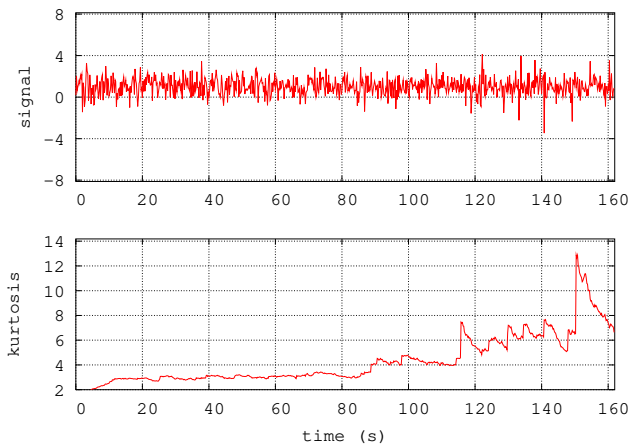,width=\textwidth}}
\end{minipage}
 \caption{\label{example1}
   \textbf{Recursive estimate of the kurtosis of an evolving mixture of Gaussian and Laplacian
     noises} {\protect \cite{02:_gnu_octav}}. \textit{top}: the test signal we use is a mixture of
   noises (both of unit mean and variance) having resp. a normal PDF ${\cal N}(1,1)$ and a Laplace
   PDF $\propto \exp(-\sqrt{2} \abs{x-1})$. The two noises are combined linearly. The weight
   coefficient of the Laplacian noise increases from 0 to 1 (and reverse for Gaussian) following a
   linear function of time.  \textit{bottom}: we see in this plot that an excess of kurtosis
   $\hat{\kappa}_4(t)$ (see Sect.  {\protect \ref{validation}} for details) appears starting from
   $t\approx 80$s (time at which the Laplacian noise starts to dominate).}
\end{figure}

\noindent \textbf{Check \#2: is the estimator useful for detecting noises with heavy tails? ---}
Figure \ref{example1} presents the result of the estimation of the kurtosis of an evolving mixture
of Gaussian and Laplacian noises. The kurtosis of a Laplace random variable is equal to $\kappa_4=6$
indicating that the tails of this distribution $\propto \exp(-\sqrt{2}\abs{x})$
\cite{johnson70:_distr_statis} decrease slowly as compared to the Gaussian ones. The two noises are
linearly combined ; the weight coefficient of the Laplacian noise increases from 0 to 1 (and reverse
for the Gaussian noise) following a linear function of time. An excess of kurtosis (i.e.,
$\hat{\kappa}_4(t) \gtrsim 4$) appears starting from $t\approx 80$s (time at which the Laplacian
noise starts to dominate) showing that we can answer positively to the question.

\begin{figure}
\begin{minipage}[b]{1.0\linewidth}
  \centering
  \centerline{\epsfig{figure=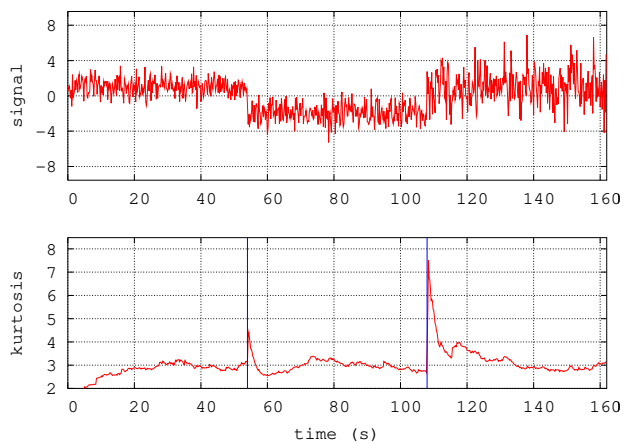,width=\textwidth}}
\end{minipage}
 \caption{\label{example2}
   \textbf{Recursive estimate of the kurtosis of a Gaussian noise of changing mean and variance}
   {\protect \cite{02:_gnu_octav}}.  \textit{top}: this is the same simulation as in Fig.  {\protect
     \ref{example1}} with a white Gaussian noise whose mean and variance are set resp. to the following
   values $(1, -2, 1)$ and $(1, 1, 4)$ during the three periods ($t<54$ s, $54 \leq t < 108$ s,
   $t>108$ s).  \textit{bottom}: after a transient period roughly equal to the window
   length, $\hat{\kappa}_4(t)$ tends to the correct value which is 3. Each change of the mean or
   variance is seen as a non-Gaussianity (i.e., large values of the kurtosis).}
\end{figure}

\noindent \textbf{Check \#3: effect of non-stationarities ---}
Figure \ref{example2} illustrates how the recursive estimator of the kurtosis performs
with a simulated Gaussian noise of changing mean and variance. After a transient period roughly
equal to the window length, $\hat{\kappa}_4(t)$ tends to the correct value which is 3. Each change
of the mean or variance is seen as a non-Gaussianity (i.e., large values of the kurtosis).  The
reason is that the hypothesis of local stationarity required for a correct estimation (see Sect.
\ref{recursive_est}) is not satisfied at the discontinuity points.

\noindent \textbf{Practical use with gravitational wave data ---}
For technical reasons related to the common data format used by the gravitational wave detectors, it
is convenient to fix the output rate of the monitoring indicators to one sample per (GPS) second of
data (also referred to as \textit{frame}). This applies to the normality index we would like to set
up. 

Since the variance of $\hat{\kappa}_4(t)$ is difficult to obtained, we cannot compute a confidence
interval which would be required to conclude on the normality of the data from the estimator value.
We remedy this with the following scheme~:
\begin{enumerate}
\item choose arbitrarily a threshold $\eta$ (e.g., $\eta=4$),
\item associate a warning flag to each frame of data where an excess of kurtosis (i.e.,
  $\hat{\kappa}_4>\eta$) has been observed at least once,
\item compute the rate of triggered frames (over periods corresponding to the typical duration of a
  GW as seen by the detector),
\item compare this rate to the one evaluated with Monte-Carlo simulations using a Gaussian noise
  with similar spectral characteristics than the signal being observed.
\end{enumerate}

Figure \ref{pdf} (bottom) gives the expected value (with error bars) of this rate if the signal is
Gaussian and white, for thresholds taken between 2.4 and 4. For instance, if we fix $\eta=4$, then
rates of triggered frames larger than 1\% indicate the presence of a heavy-tailed noise in the data.

\begin{table}[htb]
\begin{minipage}[b]{1.0\linewidth}
\begin{boxedverbatim}

choose a window duration -> define a1

C1=1-a1
C2=(1-a1^2)/2
bias=-3 C_1

init mu1_last, mu2_last, k4_bar_last

while (data available)
  x = get next data sample

  mu1= a1 mu1_last + C1 x
  dx2= (x-mu1_last)^2
  mu2= a1 mu2_last + C2 dx2
  dx2= dx2/mu2_last
  k4_bar= (1+C1-2 C1 dx2) k4_bar_last ...
          ... + C1 dx2^2
  kappa4= k4_bar + bias

  send kappa4 to output

  mu1_last= mu1
  mu2_last= mu2
  k4_bar_last=k4_bar
end

\end{boxedverbatim}
\end{minipage}
\caption{\label{code}
   \textbf{Pseudo-code for the computation of $\hat{\kappa}_4(t)$}.
The algorithm requires a total number of 16 floating point operations (10 multiplications
and 6 additions) to compute the next kurtosis value from the previous one. The memory usage is
restricted to three registers of real numbers. As a comparison, a sliding estimate using k-Statistics
{\protect \cite{kenney51:_mathem_statis}} would need at least a register of the same size than the window. 
Note that there are several possible initializations of the three registers in line 5. In our simulations, 
we set them to 0, 1 and 0 respectively. This initialization affects essentially the transient period at the 
beginning of a computation.}
\end{table}

\vfill\pagebreak

%\begin{figure}[htb]

\bibliographystyle{IEEEbib}
\bibliography{paper}

\begin{thebibliography}{1}

\bibitem{gwdetectors}
,'' Here is a list of Internet sites where more information can be found on
  respective detectors~: LIGO (\url{http://www.ligo.caltech.edu}), TAMA300
  (\url{http://tamago.mtk.nao.ac.jp}), GEO600
  (\url{http://www.geo600.uni-hannover.de}), VIRGO
  (\url{http://www.virgo.infn.it}).

\bibitem{johnson70:_distr_statis}
N.~L. Johnson and S.~Kotz,
\newblock {\em Distribution in Statistics. {D}iscrete distributions --
  {C}ontinuous Univariate distributi ons},
\newblock Wiley, New York, 1970.

\bibitem{amblard95:_adapt}
P.~O. Amblard and J.~M. Brossier,
\newblock ``Adaptive estimation of the fourth-order cumulant of a white
  stochastic process,''
\newblock {\em Signal Processing}, vol. 42, no. 1, pp. 37--43, 1995.

\bibitem{aarts02:_effic}
R.M. Aarts, R.~Irwan, and A.J.E.M. Janssen,
\newblock ``Efficient tracking of the cross-correlation coefficient,''
\newblock {\em IEEE Trans. on Speech and Audio Proc.}, vol. 10, no. 6, pp.
  391--402, 2002.

\bibitem{kenney51:_mathem_statis}
J.~F. Kenney and E.~S. Keeping,
\newblock {\em Mathematics of Statistics},
\newblock Van Nostrand, New York, 2nd edition, 1951.

\bibitem{02:_gnu_octav}
,'' All simulations and figures shown in this article were made with {GNU}
  {O}ctave and {G}nuplot. \url{http://www.octave.org}.

\end{thebibliography}

%\vfill
%\pagebreak

\end{document}